\documentclass[12pt, preprint]{elsarticle}

\usepackage{graphicx}                 
\usepackage{amsmath,amssymb}          
\usepackage[version=4]{mhchem}        
\usepackage{booktabs}                 
\usepackage{microtype}                
\usepackage[hyphens]{url}             

\usepackage[colorlinks=true,
            linkcolor=black,
            citecolor=black,
            urlcolor=black]{hyperref}
            
\begin{document}

\begin{frontmatter}

\title{Computational Design of Ductile Additively Manufactured Tungsten-Based Refractory Alloys}

\author[aff1]{Kareem Abdelmaqsoud}
\author[aff2]{Daniel Sinclair}
\author[aff2]{Venkata Satya Surya Amaranth Karra}
\author[aff2,aff3]{S.~Mohadeseh Taheri-Mousavi}
\author[aff4]{Michael Widom}
\author[aff2]{Bryan A. Webler}
\author[aff1]{John R. Kitchin\corref{cor1}}

\cortext[cor1]{Corresponding author}
\ead{jkitchin@andrew.cmu.edu}

\affiliation[aff1]{organization={Department of Chemical Engineering}, organization={Carnegie Mellon University}}
\affiliation[aff2]{organization={Department of Materials Science and Engineering}, organization={Carnegie Mellon University}}
\affiliation[aff3]{organization={Department of Mechanical Engineering}, organization={Carnegie Mellon University}}
\affiliation[aff4]{organization={Department of Physics}, organization={Carnegie Mellon University}}

\begin{abstract}
Tungsten exhibits exceptional temperature and radiation resistance, making it well-suited for applications involved in extreme environments such as nuclear fusion reactors. Additive manufacturing offers geometrical design freedom and rapid prototyping capabilities for these applications, provided the intrinsic brittleness and low printability of tungsten can be overcome. Designing tungsten alloys with improved ductility (and thus printability in AM) can be accelerated by the use of a computationally-derived performance predictor to screen out brittle compositions. Calculations of the Pugh ratio by density functional theory (DFT) may serve this purpose well, given the value's correlation with ductility. Further, this process can be made more efficient by the use of machine learning interatomic potentials (MLIPs) to accelerate DFT calculations. Here, we demonstrate that MLIPs can effectively identify optimal alloy compositions in the W–Ta–Nb system along the melting point–Pugh ratio Pareto front. The trend of the Pugh ratio as a function of tungsten fraction in the alloys is explained in terms of the electronic density of states at the Fermi level. Experimental validation reveals a strong correlation between the computed Pugh ratio and the observed crack fractions in additively manufactured alloys. Notably, the two alloys predicted to have the highest Pugh ratio values, $\ce{W20Ta70Nb10}$ and $\ce{W30Ta60Nb10}$, exhibit no intergranular micro-cracking in experiments.
\end{abstract}

\end{frontmatter}

\section{Introduction}
Recent developments in magnetically confined plasma technology have led to the construction of new Tokamak fusion reactors in the US \cite{SPARC}, China \cite{BEST}, and the European Union \cite{ITER}. The successful long-term operation of these reactors requires plasma-facing materials capable of withstanding high particle fluxes and extreme temperatures while retaining oxidation resistance, ductility, and hardness \cite{ITER}. In current fission reactors, reactor materials limit outlet temperatures to 1000$^\circ$C\cite{NUCLEAR_PLANTS}. While graphite support structures can exceed this temperature, a metallic material with an operating temperature up to 1500$^\circ$C would expand the safety margins and production capacity of new reactors. By estimating operating temperature as $0.5T{M}$ and applying a conservative safety factor of 20\%, we conclude that a material will require a melting temperature of 3300K to functionally improve reactor performance. This has historically been the motivation to develop tungsten as a reactor shielding material and for the development of tungsten-containing alloys.  

Research on multi-component/high-entropy alloys provides significant evidence that solid solutions of transition metals with high melting temperatures may yield the properties needed in future nuclear energy applications \cite{CANTOR_MCA,YEH_HEA}. Since 2010, refractory multi-principal element alloys (RMPEA's) have been explored in over 70 papers, covering 11 different alloy systems \cite{PLASMA_FACING}. Repeatedly, alloys with multiple principal components demonstrate strength and temperature resistance beyond what is predicted by the rule of mixtures due to extreme lattice distortion, which resists self-diffusion and dislocation motion while preserving ductility \cite{HEA_REVIEW}. As an example, compressive testing of equiatomic W-Ta-Nb-Mo and W-Ta-Nb-Mo-V alloys showed a peak strength over 1 GPA at room temperature and up to 500 MPa at 1600$^\circ$C \cite{SENKOV_EQUI}. However, manufacturability presents a unique obstacle to the adoption of tungsten-containing RMPEA's. Forming and cutting processes are complicated by the material's extreme hardness, and part complexity is limited. As a result, new alloy compositions and manufacturing routes are being pursued.

Increasingly, research in tungsten RMPEA's focuses on the additive manufacturing (AM) of functional features such as hollow lattices and cooling channels. However, the current literature exposes a barrier to tungsten AM: room-temperature ductility. Defect-free tungsten shows a ductile-to-brittle transition temperature between 300 and 700$^\circ$C; as a result, printed parts crack easily under the high thermal stresses of AM processing \cite{DBTT, DBTT2}. The contribution of oxidation to this process requires controlled environments, and only a narrow process window exists where solid parts can be manufactured \cite{KARRA_WTA}. The optimization of ductility in tungsten alloys is thus a necessity to enable the defect-free AM of tungsten. The combinatorial nature of RMPEA's offers ample opportunities to search for enhanced ductility but requires a strategic approach. Screening compositions using predictive computational techniques—such as machine learning \cite{ALLOYGPT, guduru_ductgpt_2026} or electronic structure simulation via density functional theory (DFT) \cite{sholl_density_2022, ouyang_design_2023, ELECTRONICDUCTILITY}—enables efficient navigation of this vast design space. Additionally, alloys can now be tested more quickly using powder-blown directed energy deposition (PB-DED) or laser engineered net shaping (LENS), giving rise to a framework of high-throughput alloy design\cite{THROUGHPUT}. A large body of work has focused on optimizing high-temperature strength using this approach, but we propose that optimizing printability is a more crucial first step in the design of a tungsten-based RMPEA. An empirical connection is thus needed between cracking in AM tungsten alloys and a theoretical ductility criterion.

DFT is uniquely well-suited to estimate the ductility of RMPEAS and yields multiple property measurements which may correlate well to physical behavior \cite{ouyang_design_2023, shaikh_designing_2023}. The Pugh ratio, defined as the ratio of the bulk modulus ($K$) to the shear modulus ($G$), has been correlated with ductility \cite{pugh_xcii_1954}. A higher Pugh ratio $K/G$ generally corresponds to more ductile behavior, although its alignment with the printability of RMPEA's has not been demonstrated experimentally. This may be due to the defect-rich microstructures that are common to AM parts, which degrade material ductility to the point that mechanical testing cannot accurately describe plastic deformation. We posit that intergranular microcracking during printing may provide a quantifiable feature that can be isolated relative to other process-dependent effects. In DED parts, the cross-section of a single deposited track may include hundreds of grains, which experience thermal stresses as a global force. Provided processing parameters like laser power and speed are held constant, grains should experience a consistent stress, which will be accommodated first by lattice deformation and second by grain boundary cracking. By quantifying the densities of microcracks in alloyed samples, we seek to conclusively report a correlation (or lack thereof) to the Pugh ratio and support the use of ab initio ductility predictions, which have been widely adopted but rarely verified by experimental data. 

The second feature of DFT alloy screening that this work seeks to address is the high computational cost of predicting properties in large combinatorial spaces. This is done by applying machine learning interatomic potentials (MLIPs). MLIPs have emerged as accurate and computationally efficient surrogates for DFT that enable the prediction of atomic energies, forces, and elastic properties at orders of magnitude lower computational cost. MLIPs are typically evaluated on errors in the energy and force predictions; however, these metrics do not always translate to more accurate physical property predictions. Herein, we evaluate their accuracy in calculating elastic properties, such as shear and bulk moduli, in relation to the ground truths of DFT and experimental results.

Our present verification of DFT and MLIP predictions is focused on the W-Ta-Nb alloy system, which is representative of AM tungsten alloys that may achieve suitable ductility for printing with minimal loss of high-temperature properties. We use the Universal Model for Atoms (UMA) potential, a unified model capable of accurately reproducing DFT-level predictions across different material classes without system-specific fine-tuning \cite{wood_uma_2025, barroso-luque_open_2024, levine_open_2025, chanussot_open_2021, sriram_open_2025, gharakhanyan_open_2025}. We use the UMA potential to systematically calculate the Pugh ratios of W–Ta–Nb alloys across the ternary composition space. The Pugh ratios of the alloys that lie on the Pareto front of the Pugh ratio and the melting point were validated using DFT. The electronic density of states at the Fermi level was used to explain the trend in Pugh ratio as a function of the tungsten fraction in the alloys  \cite{pant_electronic_2025}. Experimental validation reveals a strong correlation between the computed Pugh ratio and the observed crack fractions in additively manufactured alloys. Notably, the two alloys predicted to have the highest Pugh ratio values, \ce{W20Ta70Nb10} and \ce{W30Ta60Nb10}, exhibit no intergranular cracking in experiments. These results indicate a promising pathway to optimize alloy printability in future RMPEA development.

\section{Methods}

\subsection{Computational Setup}

Given that all alloys in the W–Ta–Nb system exhibit solid-solution behavior, their lattices were modeled using special quasi-random structures (SQS) \cite{zunger_special_1990}, which approximate the atomic disorder of random alloys. The SQS configurations were generated using the ICET package \cite{angqvist_icet_2019}. All possible W–Ta–Nb compositions containing ten atoms were enumerated, and SQS structures were generated for each composition to systematically sample the ternary compositional space. Figure S7 shows that using SQS structure with a 10-atom size is sufficient, as the Pugh ratio does not change significantly when a larger size of 50 atoms is used.

The UMA MLIP was chosen compared to other MLIPs, such as MACE-OMAT \cite{batatia_mace_2023}, because it accurately predicts the DFT Pugh ratios for the pure elements with a small overestimation of the Pugh ratio of Ta. Unlike MACE-OMAT, the UMA model captures the correct trend in ductility of the pure elements Nb$>$Ta$>$W as shown in Figure S1. To calculate the bulk and shear moduli, all structures were first relaxed using the OMAT head of the UMA model and then subjected to a series of normal and shear deformations to evaluate the elastic tensor. Linear strain values of (-1\%, -0.5\%, 0.5\%, 1\%) and (-6\%, -3\%, 3\%, 6\%) were applied for the normal and shear modes, respectively. Since shear deformations generally induce a weaker elastic response than normal deformations, larger strain amplitudes are commonly employed in the literature to ensure an accurate stress–strain fit for MLIPs\cite{kaplan_foundational_2025}. As shown in Figure S4 of the Supplemental Materials, the stress–strain relationship remains linear at these shear strain magnitudes. All deformations and elastic tensor fittings were performed using the Pymatgen package\cite{ong_python_2013}.

The bulk ($K_{\mathrm{VRH}}$) and shear ($G_{\mathrm{VRH}}$) moduli were calculated from the elastic tensor using the Voigt–Reuss–Hill (VRH) averaging scheme \cite{hill_elastic_1952}. The following equations were used to calculate the bulk and shear moduli for the BCC alloys considered in this work:

\begin{align}
K_{\mathrm{VRH}} &= \frac{C_{11} + 2C_{12}}{3}, \label{eq:K_BCC} \\
G_V &= \frac{C_{11} - C_{12} + 3 C_{44}}{5}, \label{eq:GV_BCC} \\
G_R &= \frac{5}{4 s_{11} - 4 s_{12} + 3 s_{44}}, \label{eq:GR_BCC} \\
G_{\mathrm{VRH}} &= \frac{1}{2} \left( G_V + G_R \right), \label{eq:GVRH_BCC}
\end{align}

where $s_{ij} = C_{ij}^{-1}$ is the compliance matrix and $C_{ij}$ is the elastic tensor. The Pugh ratios ($K_{\mathrm{VRH}}/G_{\mathrm{VRH}}$) of the alloy compositions were calculated from the corresponding bulk and shear moduli.

Our analysis focuses not only on alloys with high Pugh ratios, indicative of enhanced ductility, but also on alloys with high melting points, specifically those exceeding 3300K. The 3300K melting point criterion was chosen to target high working temperature alloys, around 1500°C. Melting points were estimated using Vegard’s law, based on a linear combination of the melting points of the constituent pure elements \cite{denton_vegards_1991, chelikowsky_melting_1987}. A Pareto front of Pugh ratios versus melting points was then constructed to identify compositions with optimal performance in both properties. Alloys lying on the Pareto front and meeting the 3300K melting point criterion were further investigated using DFT to confirm the trends with higher accuracy.

The DFT calculations were performed using the Vienna Ab initio Simulation Package (VASP) with the PBE functional \cite{kresse_efficiency_1996,kresse_ultrasoft_1999,perdew_generalized_1996}. A $k$-point grid density of $7000 / N_\text{atoms}$ was employed to ensure accurate energy calculations for the structures under the small applied deformations. A high plane-wave energy cutoff of 520 eV was used. The recommended VASP pseudopotentials were employed: \texttt{W\_sv}, \texttt{Ta\_pv}, and \texttt{Nb\_pv}. All structures were tightly relaxed (EDIFF = $1\times10^{-7}$) prior to applying deformations using the finite-difference method (IBRION = 6, ISIF = 3).

\begin{figure}[h]
    \centering
    \includegraphics[width=0.9\textwidth]{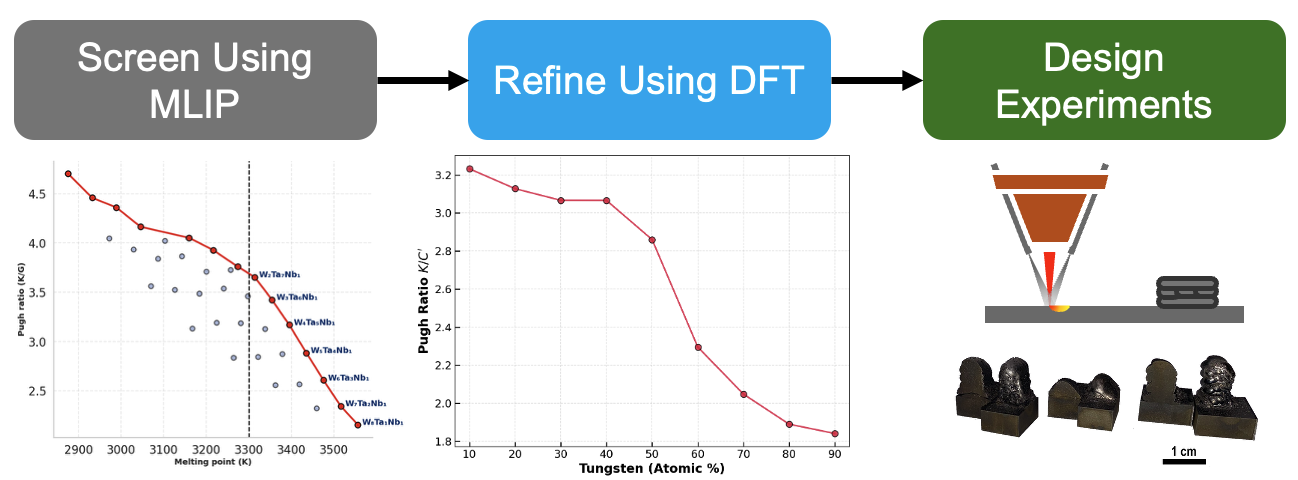}
    \caption{Pipeline for using machine learning interatomic potentials (MLIPs) to screen the ternary alloy space, obtain a more accurate trend using DFT, and use that to design experiments of ductile high-temperature W-Ta-Nb alloys.}
    \label{fig:toc}
\end{figure}

Figure \ref{fig:toc} shows the full workflow for using MLIPs and DFT to guide the experimental design of W-Ta-Nb alloys. The MLIP is used to screen the ternary alloy space, obtain a more accurate trend using DFT, and use that to design experiments of ductile high-temperature W-Ta-Nb alloys.

\subsection{Experimental Setup}

\begin{figure}[h]
    \centering
    \includegraphics[width=0.8\textwidth]{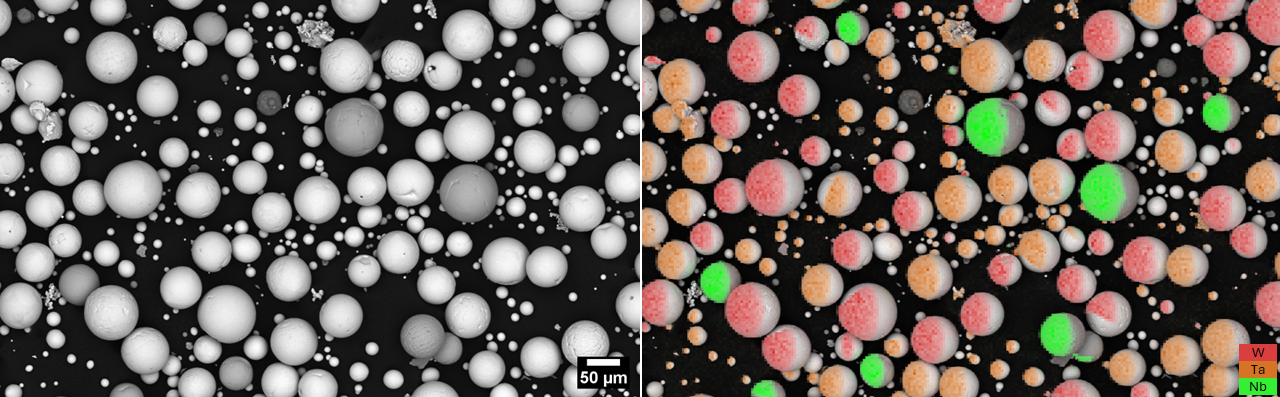}
    \caption{SEM/EDS images of blended powder with nominal atomic composition W50-Ta40-Nb10}
    \label{fig:Powders}
\end{figure}

\begin{figure}[t]
    \centering
    \includegraphics[width=0.8\textwidth]{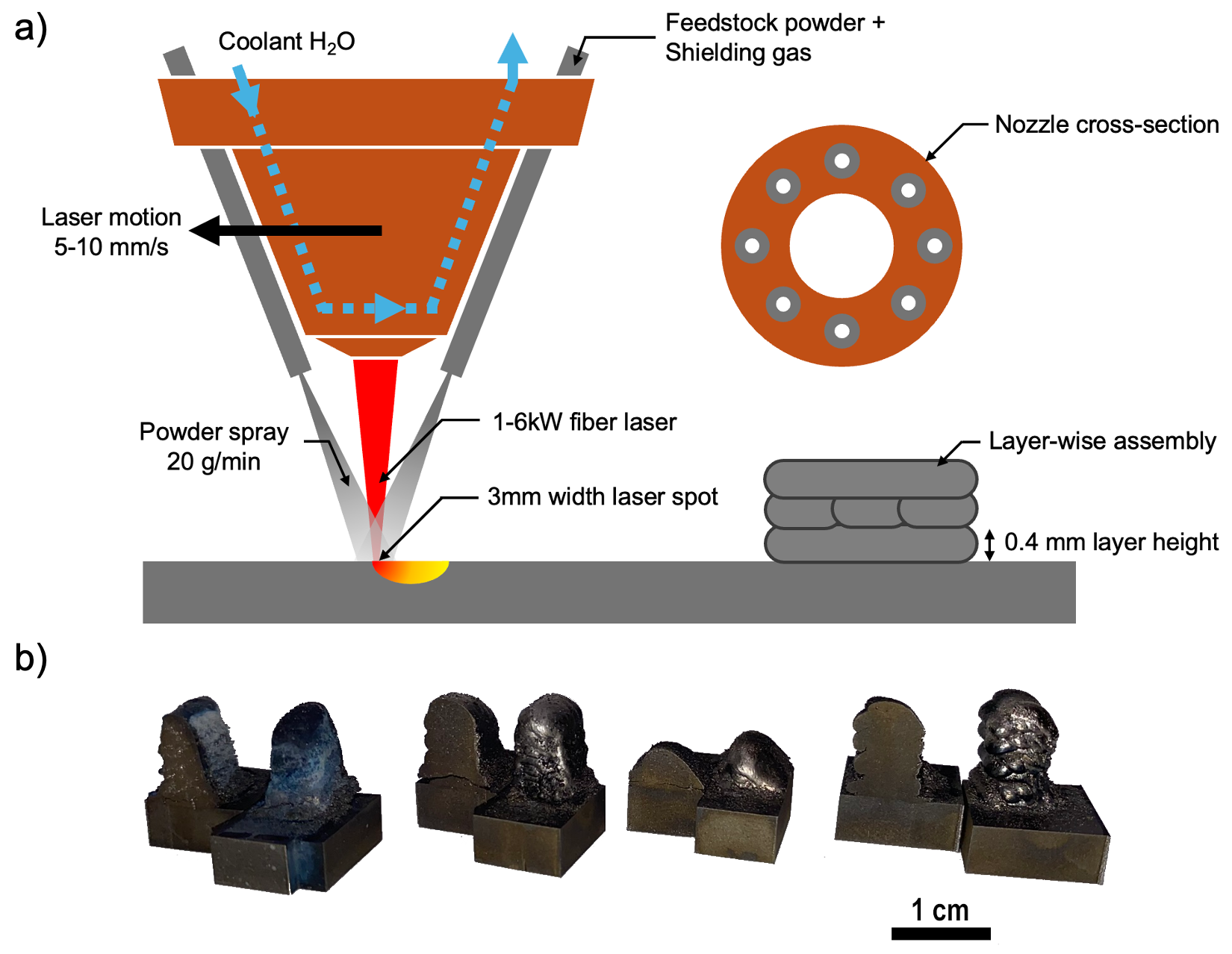}
    \caption{a) Schematic of DED nozzle operation during deposition of 3D walls, b) examples of cross-sectioned walls demonstrating the range of geometries and surface finishes produced by DED. }
    \label{fig:am_alloys}
\end{figure}

Alloys were prepared from plasma-spheroidized tungsten, tantalum, and niobium powders having purity $>99.98\%$ (purchased from Heeger Materials, Denver, CO). All powders had a size range of 45-106 $\mu$m and were blended manually in a clean vessel (Figure \ref{fig:Powders}). Blended powders were then deposited using a Trumpf TruLaser3000 laser cutting/deposition system equipped with an 8-nozzle MultiJet nozzle. Figure \ref{fig:am_alloys}a shows a schematic of DED nozzle operation during deposition of 3D walls. Seven alloys were produced, with tungsten content ranging from 20-80 atomic $\%$ and an average niobium content of 11.5$\pm$1.3 atomic $\%$. The relative weight fraction of each constituent powder was calculated to achieve tungsten atomic fractions of 20, 30, 40, 50, 55, 60, and 80; however, the compositions of printed parts varied due to powder settling, deposition effects, and experimental error. Thus, one alloy was removed from consideration; the measured compositions of the six analyzed alloys are presented in Table \ref{tab:compositions}. All compositions will be discussed in atomic fractions to maintain consistency with DFT measurements.

\begin{table}
    \centering
    \caption{Atomic fraction (at.\%) of principal elements in measured samples, recorded by EDS ($n=16$)}
    \label{tab:compositions}
    \begin{tabular}{l c c c} 
        \toprule
        Sample & \multicolumn{3}{c}{Atomic Fraction (at.\%)} \\
        \cmidrule(lr){2-4} 
        ID & W & Ta & Nb \\
        \midrule
        1 & $80 \pm 3$ & $8 \pm 1$ & $12 \pm 3$ \\
        2 & $64 \pm 2$ & $26 \pm 2$ & $10 \pm 3$ \\
        3 & $56 \pm 4$ & $32 \pm 2$ & $12 \pm 4$ \\
        4 & $51 \pm 3$ & $37 \pm 2$ & $13 \pm 4$ \\
        5 & $30 \pm 1$ & $57 \pm 1$ & $13 \pm 1$ \\
        6 & $23 \pm 1$ & $67 \pm 1$ & $10 \pm 1$ \\
        \bottomrule
    \end{tabular}
\end{table}

Depositing tungsten requires a low-oxygen environment; thus, builds were conducted inside an argon-purged chamber constructed from a flexible aluminized fiberglass wall, a steel baseplate with a silicone seal, and an aluminum plate sealing the wall to the nozzle. The chamber was purged with argon until an oxygen content of $<10$ ppm was achieved. This approach is capable of producing tungsten parts with oxygen content between 11-37 ppm\cite{KARRA_WTA}.

During deposition, laser power, speed, and spot size were kept constant at 4750 W, 300 mm/min, and 3 mm, respectively. The powder mass deposition rate was 0.3 g/s and maintained by adjusting the amplitude of vibrations in the powder hopper before each build. A niobium build plate was used without preheating. Samples were constructed by alternating four parallel 15 mm tracks with ten 6 mm tracks rotated by 90$^\circ$. Track spacing and vertical offset were based on the cross-sections of single tracks; track spacing was set equal to half of the cross-sectional area divided by width, and vertical offset was equal to the track height. Examples of the resulting geometries are shown in Figure \ref{fig:am_alloys}b.

After building, samples were cross-sectioned using electron discharge machining, mounted in two-part epoxy, and ground until flat. They were then polished using silicon carbide paper, diamond suspensions, and finally colloidal alumina with particle size 0.5 $\mu$m. For two samples (4 and 6 in Table \ref{tab:compositions}), the opposite face of the sectioned wall was mounted, polished, and etched with Murakami's reagent (10 g potassium ferricyanide, 10 g potassium hydroxide, and 100 mL water) by immersing for 30 seconds.

\begin{figure}[t]
    \centering
    \includegraphics[width=0.6\textwidth]{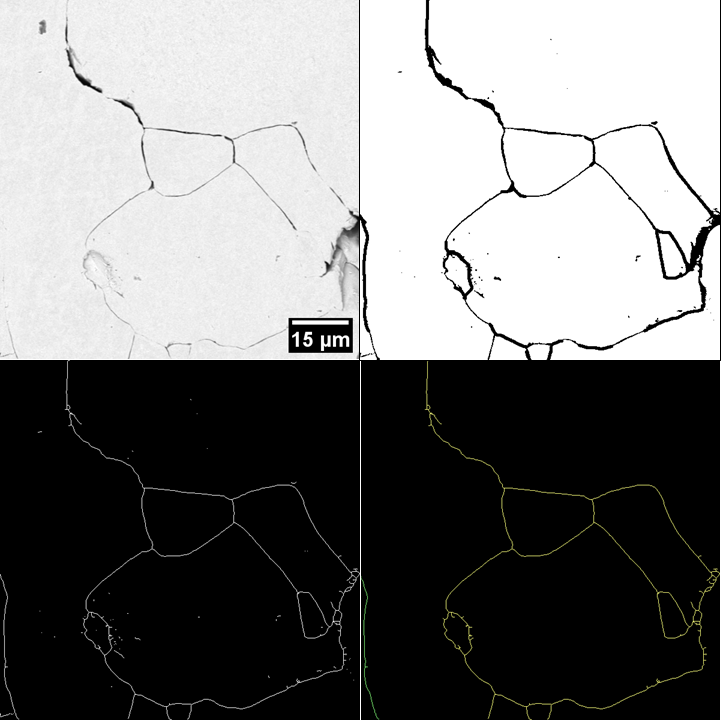}
    \caption{(Left to right, top to bottom) SEM image of microcracking in tungsten alloy, followed by manual binarization, skeletonization, and labeling}
    \label{fig:CrackSegmentation}
\end{figure}

To measure microcracking, polished samples were imaged in a ThermoScientific ParticleX desktop microscope in the backscatter electron mode with an accelerating voltage of 10 kV. Rectangular areas with areas of at least 4 mm$^2$ were imaged by combining a grid of micrographs with a pixel size below 0.5 $\mu$m. Working in ImageJ, dark regions of the stitched images were selected by a threshold tool, including cracks, pores, and surface debris. Unwanted features were manually removed, and cracks were manually corrected by comparing to the grayscale image. Where grains had fallen out fully, the outermost boundary of the pit was selected to record only features that were in the plane of the image. Cracks were then skeletonized, giving them a width of one pixel, and features with a size less than 15 pixels were removed. Finally, the total crack length was computed using a skeleton analysis Python library\cite{SKAN}, and the crack length was compared to image area to provide a crack density measurement with units mm/mm$^2$, which is directly comparable to the density of two-dimensional crack planes in the three-dimensional sample volume. Figure \ref{fig:CrackSegmentation} shows the processing steps of quantifying the cracking in an image of the alloy.

All crack measurements were made in the uppermost layer of the deposited walls (between 6-8 mm) to avoid contamination by the build plate, and areas with clear signs of cracking due to lack of fusion (e.g. delamination between deposited tracks) were intentionally avoided. Inside each analysis area, composition was measured by energy dispersive spectroscopy (EDS). Sixteen point scans were performed with 20 kV accelerating voltage, and the atomic content of tungsten, niobium, and tantalum are reported with their standard deviations in Table \ref{tab:compositions}.

The high-resolution analysis of cracks over a relatively small area imposes limitations that are known to the authors. Our measurements are indicative only of microcracking between grains, as other defects are intentionally ignored to remove the stochastic effects of DED on part density. While a broader analysis might indicate the overall quality of parts, our interest lies in a ranking of intergranular cracking in relation to alloy composition, and the most accurate comparison can be made by isolating similar regions across all conditions.

\section{Results and Discussion}

To contextualize a correlation of the Pugh ratio and ductility, the results of MLIP and DFT predictions are presented and compared, followed by an assessment of the microstructure of printed alloy specimens. The alignment of empirical and theoretical results is then discussed, along with the significance of an \textit{ab initio} printability predictor.

\subsection{Computational Results}

\subsubsection{Pugh Ratio - Melting Point Pareto-front}

\noindent

\begin{figure}[h]
    \centering
    \includegraphics[width=0.9\textwidth]{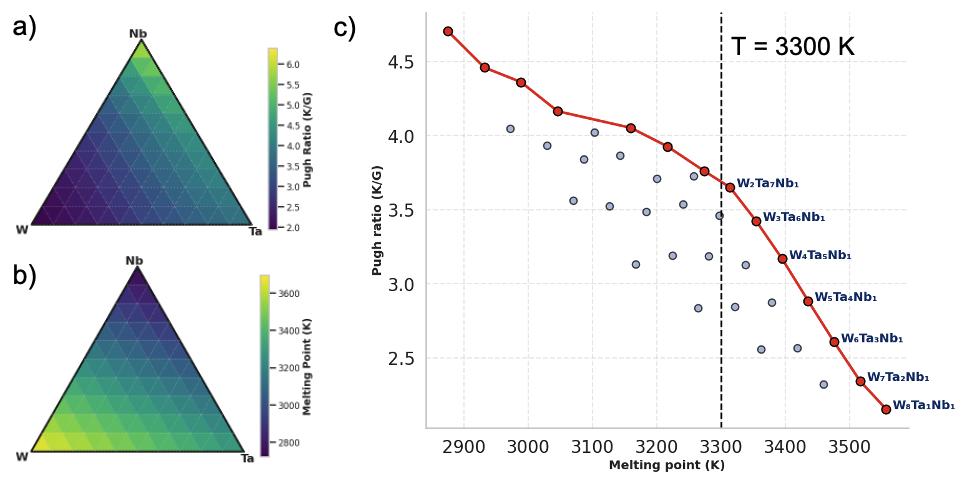}
    \caption{a) contour plots of the Pugh ratio calculated using the UMA machine learning potential. b) melting point temperatures calculated using Vegard's law. c) Pareto front of melting point temperature and Pugh ratio as a measure of ductility. A melting point of 3300 K was selected as a lower limit.}
    \label{fig:pareto_front_combined}
\end{figure}

As shown in Figures \ref{fig:pareto_front_combined}a and \ref{fig:pareto_front_combined}b, there exists a trade-off between the Pugh ratio—which is correlated with ductility—and the melting point of the alloy. When these two quantities are plotted together (Figure \ref{fig:pareto_front_combined}c), the optimal compositions form a Pareto front (highlighted in red), representing alloys that locally maximize both properties. To define a practically relevant design window, a minimum melting point of 3300 K was chosen based on the requirements of the proposed application (as discussed above). All alloys lying on the Pareto front and satisfying this melting point threshold contain approximately 10 atomic \%Nb. This trend is consistent with expectations since Nb has a lower melting point than W or Ta. Thus, a low Nb content will ideally enhance the ductility of a W-Ta-Nb alloy while minimally decreasing the melting temperature.

\subsubsection{DFT Investigation of Identified Alloys}

\noindent

\begin{figure}[h]
    \centering
    \includegraphics[width=1\textwidth]{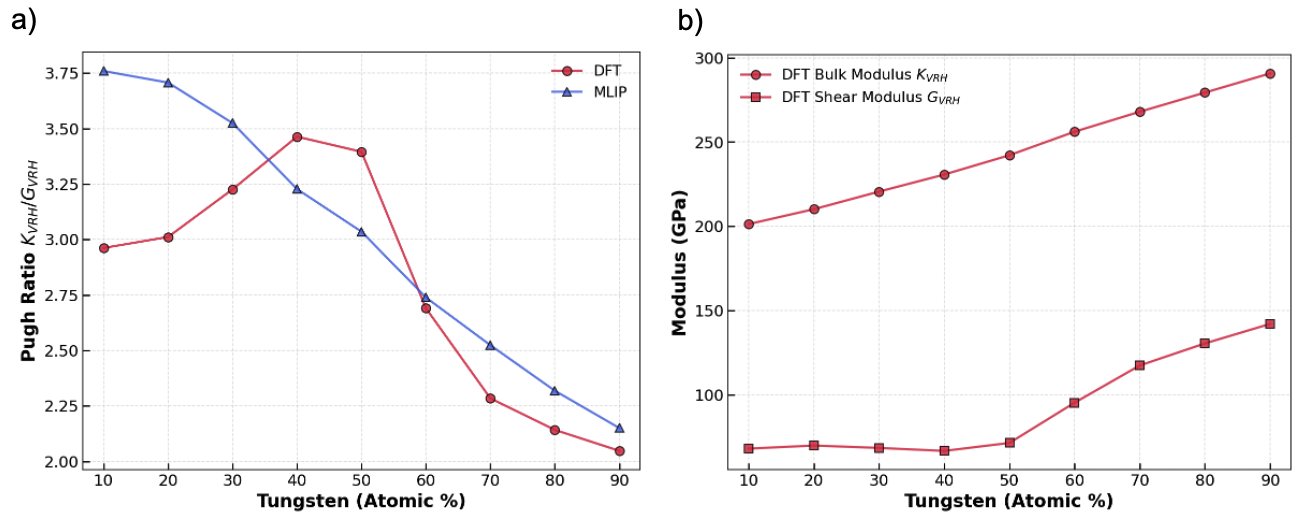}
    \caption{a)The UMA MLIP shows a monotonic decrease in the Pugh ratio with increasing tungsten atomic percentage in $\text{W}_{x}\text{Ta}_{90-x}\text{Nb}_{10}$ alloys. On the other hand, DFT results reveal a non-monotonic trend. b) The bulk modulus increases linearly as the tungsten percentage increases, whereas the shear modulus shows a plateau up to 50\% before increasing.}
    \label{fig:pugh_dft_mlp}
\end{figure}

Density functional theory (DFT) calculations were employed to further study the relationship between tungsten content and the Pugh ratio of the alloys with 10\%Nb (Figure \ref{fig:pugh_dft_mlp}a). The results from the UMA MLIP show a monotonic decrease in the Pugh ratio with increasing tungsten atomic percentage. This is expected because increasing the tungsten atomic percentage lowers ductility, which is correlated with a lower Pugh ratio. On the other hand, DFT results reveal a non-monotonic trend. The Pugh ratio increases with increasing the tungsten atomic percentage up to 40-50\% and starts decreasing at higher tungsten percentages. The DFT-derived softening behavior (higher Pugh ratio) near 40-50\% tungsten is consistent with previous observations in the binary W–Ta alloy system \cite{liu_short-range_2021}. This suggests that a similar effect persists in the ternary W–Ta–Nb alloy when Nb content is low. A deeper exploration of DFT softening behavior is needed to select the most appropriate physical model for modeling cracking.

\begin{figure}[h]
    \centering
    \includegraphics[width=1\textwidth]{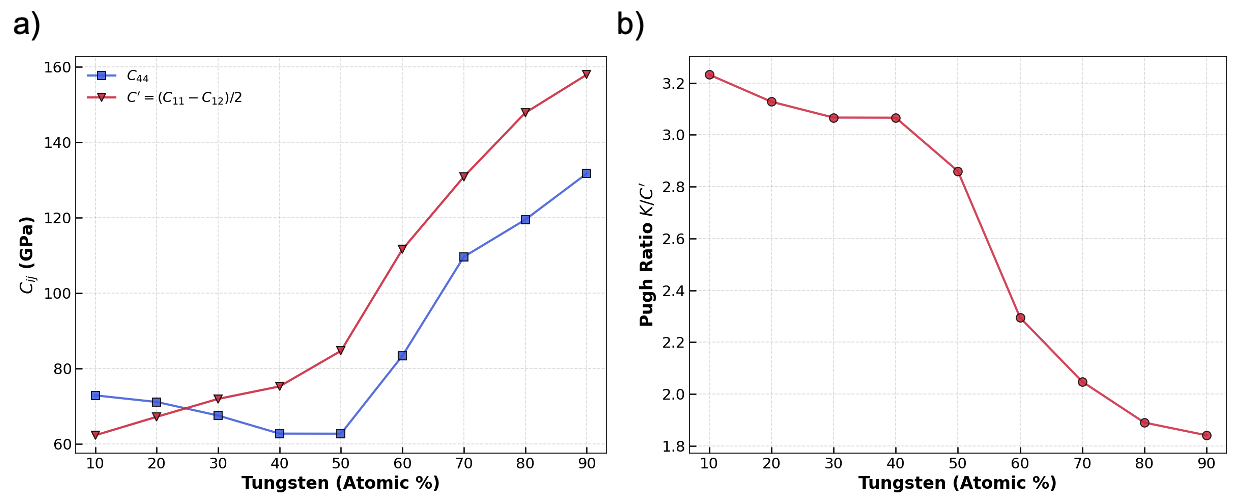}
    \caption{a) The trend of the $C_{44}$ and $C^{\prime}$ shear modes as a function of tungsten atomic percentage. For alloys containing less than $\text{50\%}$ tungsten, these two shear modes diverge. b) Pugh ratio ($K/C^{\prime}$) calculated using the tetragonal shear.}
    \label{fig:shear_modes_sqs}
\end{figure}

The origin of a softening behavior in the W-Ta-Nb alloys can be understood by decomposing the elastic response into its bulk and shear contributions (Figure \ref{fig:pugh_dft_mlp}b). While the bulk modulus increases proportionally with tungsten content, the overall shear modulus ($G_{\text{VRH}}$) exhibits a distinct plateau for tungsten percentages below $\text{50\%}$. This non-linear behavior is explained by considering the competitive and anisotropic nature of the two independent body-centered cubic (BCC) shear modes: $C_{44}$ and $C^{\prime}$. As shown in Figure \ref{fig:shear_modes_sqs}a, for alloys containing less than $\text{50\%}$ tungsten, these two shear modes diverge. The tetragonal shear ($C^{\prime}$) continues to decrease, while the $C_{44}$ shear reverses its trend and increases. These two opposite trends cause the average ($G_{\text{VRH}}$) shear modulus to plateau. This plateauing of the shear modulus is the reason for the softening of the Pugh ratio.

To identify the shear mode most relevant to cracking behavior, it is important to note that intrinsic failure in BCC metals results from competition between plastic shear and brittle cleavage under tensile loading. The $C_{44}$ shear is often used to describe dislocation-mediated plasticity and ductile failure. However, tungsten is intrinsically brittle, with failure dominated by cleavage. Qi et al. \cite{qi_tuning_2014} demonstrated that tungsten’s brittleness arises from shear instability occurring after reaching its ideal tensile strength along the tetragonal deformation path. Consequently, the stiffness constant associated with this cleavage path, the tetragonal shear modulus ($C^{\prime}$), is the more relevant descriptor for fracture resistance. Because our experimental observable is the crack fraction, which reflects brittle failure, we correlate the measured cracking tendency with the Pugh ratio defined using the tetragonal shear modulus, rather than the conventional form using $G_{\text{VRH}}$. The Pugh ratio ($K_{\text{VRH}}/C^{\prime}$) is plotted as a function of the tungsten fraction in Figure \ref{fig:shear_modes_sqs}b.

\subsubsection{Electronic Structure Explanation}

\noindent

\begin{figure}[h]
    \centering
    \includegraphics[width=1\textwidth]{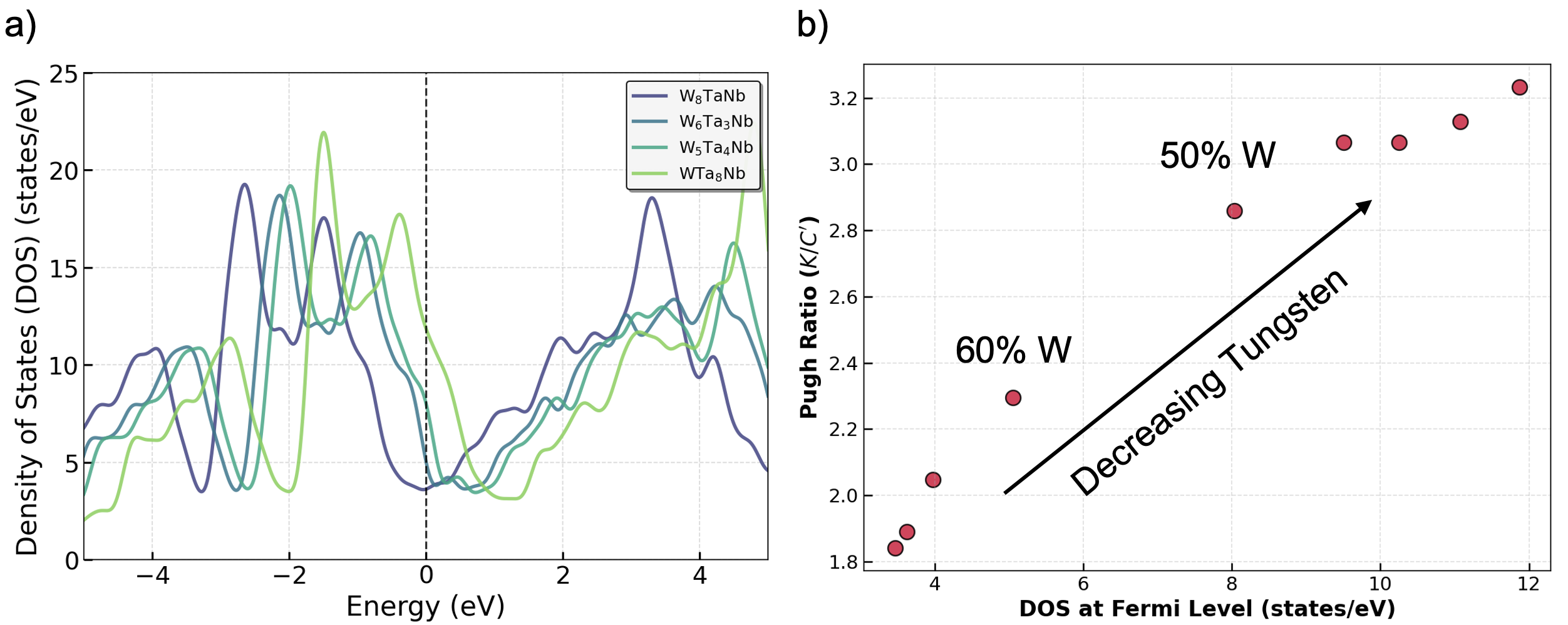}
    \caption{a) Electronic density of states plot for the different alloy compositions. The pseudogap of the alloys that have higher than 50\% tungsten overlaps with the Fermi level. b) Relationship between the Pugh ratio and the electronic density of states at the Fermi level (DOS) of W-Ta-Nb alloys with decreasing W fraction.}
    \label{fig:edos_pugh_ratio}
\end{figure}

Figure \ref{fig:shear_modes_sqs}b shows a significant drop in the Pugh ratio between 50\% and 60\% tungsten atomic percentage. The electronic density of states (DOS) at the Fermi level provides a mechanistic explanation for this trend in Pugh ratio\cite{pant_electronic_2025}. A pseudogap is a partial depletion of the electronic density of states (DOS) near the Fermi level, leading to larger energy penalties during elastic deformation and lower ductility. Pure W exhibits a pseudogap that coincides with the Fermi level, whereas Ta and Nb have their pseudogaps located above it. Alloying W with Ta and Nb shifts the pseudogap upward, increasing the density of states at the Fermi level and increasing ductility. As shown in Figure \ref{fig:edos_pugh_ratio}a, for alloys with tungsten content higher than 50\% (\ce{W80Ta10Nb10}), the Fermi level lies within the pseudogap, resulting in a low DOS. On the other hand, alloys with 50\% or lower tungsten percentage (\ce{W50Ta40Nb10}, \ce{W10Ta80Nb10}) have their pseudogap above the Fermi level, indicating a high DOS at the Fermi level. Figure \ref{fig:edos_pugh_ratio}b shows the relationship between the Pugh ratio and the electronic density of states (DOS) at the Fermi level of W-Ta-Nb alloys with decreasing tungsten fraction. As the tungsten fraction decreases, DOS at the Fermi level increases, and the Pugh ratio increases, indicating higher ductility.

\begin{figure}[h]
    \centering
    \includegraphics[width=1\textwidth]{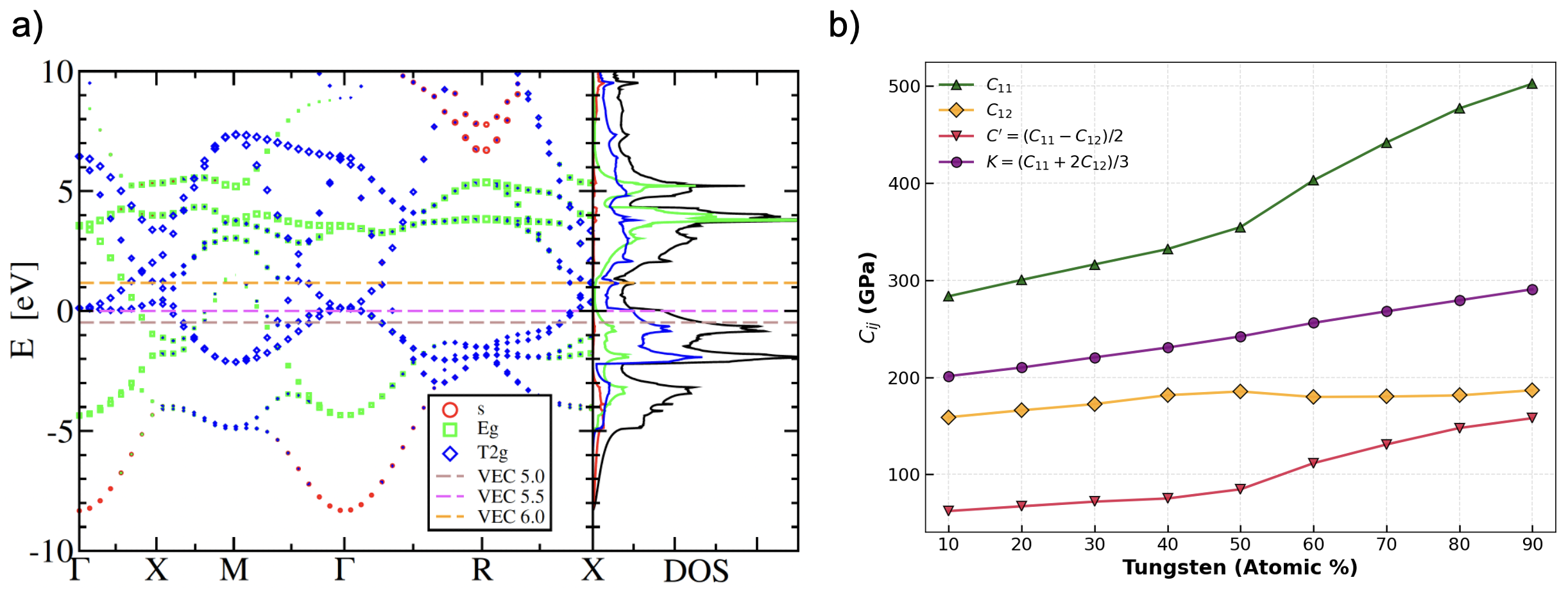}
    \caption{a) Orbital-projected electronic band structure and density of states of TaW. Marker sizes indicate the projections onto atomic orbitals. Dashed lines indicate the positions of the Fermi levels at various valence electron counts within a rigid band approximation. b) elastic components of the tetragonal shear ($C^{\prime}$) as a function of the tungsten atomic percentages.}
    \label{fig:band_structure}
\end{figure}

The drop in Pugh ratio is caused by a jump in  $C^{\prime}$ between 50\% and 60\% as shown in Figure \ref{fig:shear_modes_sqs}a. To fully understand the electronic structure origin of the jump in $C^{\prime}$, we consider the band structure and density of states (DOS) of the equiatomic binary TaW in the B2 structure. We shift the Fermi level $\text{E}_\text{F}$ to 0 and indicate the Fermi levels for elemental Ta (valence 5) and W (valence 6) within a rigid band model for the solid solution Ta$_{1-x}$W$_x$. Based on the orbital-projected DOS shown in Figure \ref{fig:band_structure}a, the $\text{T}_{\text{2g}}$ orbitals dominate around $\text{E}_\text{F}$ within the valence electron count (VEC) range of $5-6$ per atom. The three $\text{T}_{\text{2g}}$ orbitals coalesce at the $x=1/2$ Fermi level at the $\Gamma$ $k$-point. Two of the bands are nearly flat along $\Gamma-X$, while the other band is dispersive and increasing in energy. The flat band orbitals create bonds aligned in the direction of nearest neighbors, contributing to both $C_{11}$ and $C_{12}$. The dispersive orbital, in contrast, contributes primarily to $C_{11}$. As a result, $C_{12}$ grows from $x=0$ to $1/2$ and then flattens beyond $x=1/2$. Smooth growth of the bulk modulus $K=(C_{11}+2C_{12})/3$ then requires an accelerated growth of $C_{11}$, as is visible in Figure \ref{fig:band_structure}b. This result provides an electronic structure explanation for the observed trend in the Pugh ratio.

\subsection{Experimental Results}

By screening for a subset of W-Ta-Nb compositions with desirable properties (\ref{fig:pareto_front_combined}), a subset of alloys was selected to experimentally produce via PB-DED. Observations of microstructural features in six alloys are discussed here, particularly in relation to crack formation.

\subsubsection{Microstructure}

\noindent

\begin{figure}[h]
    \centering
    \includegraphics[width=0.7\textwidth]{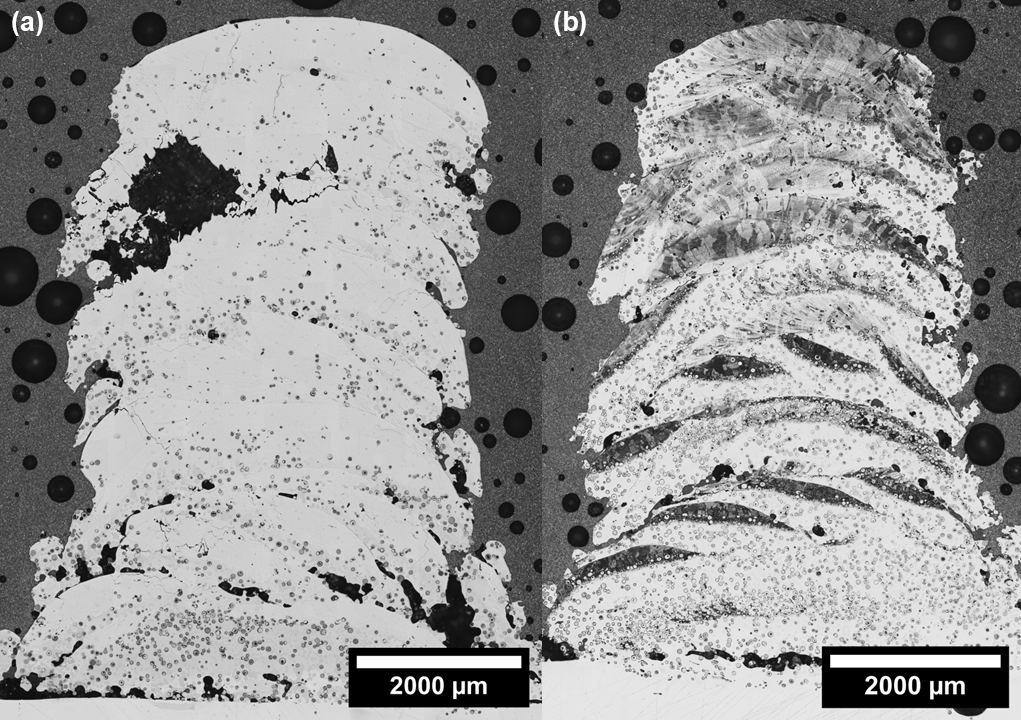}
    \caption{Cross-sections of (a) \ce{W23Ta67Nb10} and (b) \ce{W51Ta37Nb13} walls after etching with Murakami's reagent. Unmelted particles are more common in the lower layers of walls and in the W50 alloy. Regions with visible grain structure have greater than 40 percent tungsten, showing that the uppermost regions achieved the greatest homogeneity due to heat accumulation.}
    \label{fig:CrossSectionOptical}
\end{figure}

The microstructures of deposited specimens demonstrated significant heterogeneity due to the variable thermal conditions imposed by DED. Depositing on a room temperature plate led to low melt-in; thus, delamination from the plate was observed (Figure \ref{fig:CrossSectionOptical}). In the first few layers, inter-track melt-in was additionally low, leading to a lack of fusion. No solidification cracking was observed due to the low laser speed (500 mm/min) and large spot size (3 mm) used in DED. Thus, the primary process-induced defect was a lack of fusion, which may be resolved in future work through an optimization of the build plate temperature and print scheduling.

\begin{figure}[h]
    \centering
    \includegraphics[width=0.5\textwidth]{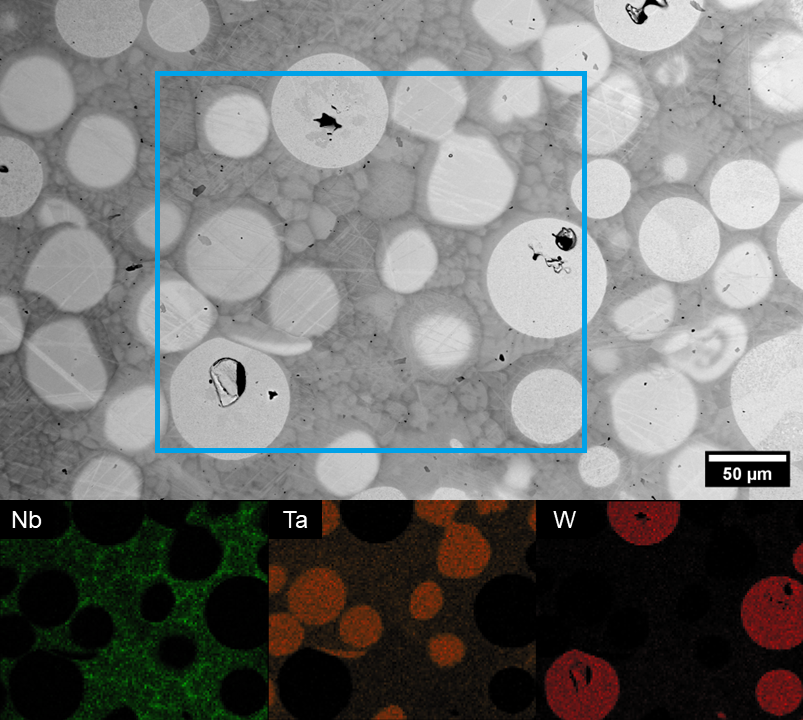}
    \caption{Unmelted tungsten and tantalum particles in a niobium-rich matrix ~1 mm from the specimen base and EDS maps of principal elements inside the highlighted scan region}
    \label{fig:UnmeltedParticles}
\end{figure}

The lowest layers of each specimen demonstrated a significant population of unmelted particles due to the significant difference in melting point between tungsten and niobium. Figure \ref{fig:CrossSectionOptical} demonstrates that a sample with higher tungsten content (23 vs. 51 atomic $\%$) has a visibly greater number of unmelted particles. EDS mapping confirms that these particles are tungsten and tantalum (Figure \ref{fig:UnmeltedParticles}). The matrix phase in this image area consisted of 51 atomic $\%$ niobium, while the niobium content in the entire region was 31$\%$ (while 10$\%$ was targeted). The issue of non-melting tungsten during DED has been demonstrated in a previous study\cite{KARRA_PARTICLES} where it was found that the accumulation of heat increased the dissolution of tungsten into the molten pool during deposition. Presently, tungsten dissolution is seen to progressively increase with each layer of the build as temperature rises. In Figure \ref{fig:CrossSectionOptical}b, etching with Murakami's reagent has revealed regions with high tungsten content (greater than 40 atomic $\%$). Higher tungsten is visible in the upper centers of melt tracks, and in the topmost deposited layer, the matrix was etched uniformly. As reported in Table 1, the composition of this region was near the targeted composition of \ce{W50Ta40Nb10}. The compositional uniformity and reduced defect population of the top two tracks support the analysis of cracking in this area. 

\subsubsection{Cracking}

\noindent

\begin{figure}[h]
    \centering
    \includegraphics[width=0.5\textwidth]{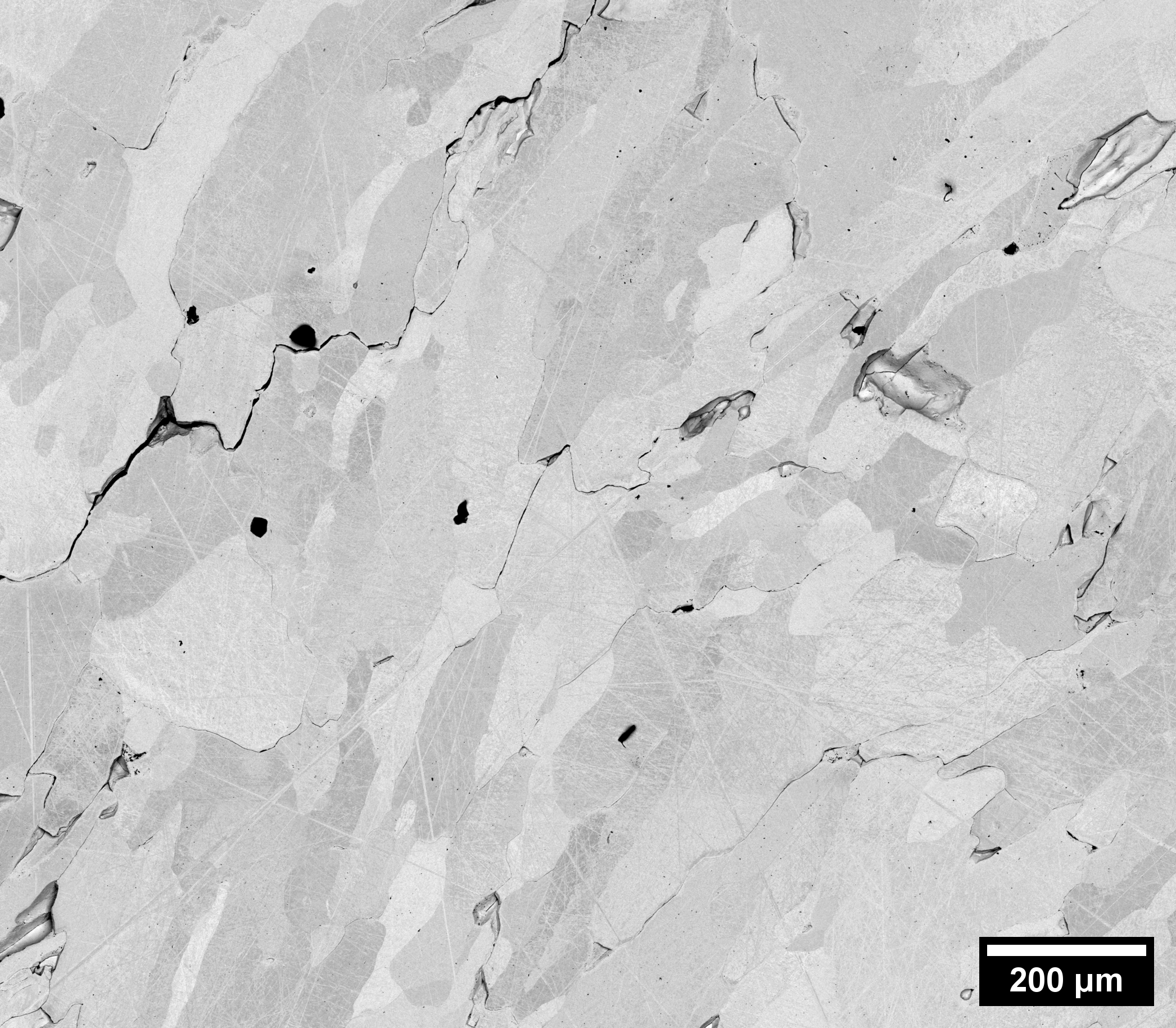}
    \caption{Polished section of \ce{W64Ta26Nb10} alloy demonstrating a network of microcracks along grain boundaries}
    \label{fig:CrackingExample-1}
\end{figure}

Throughout the microstructures of samples with more than 30 atomic $\%$ tungsten, intergranular micro-cracking was ubiquitous. BSE micrographs of areas without lack of fusion defects showed cracking between both columnar and equiaxed grains as well as an absence of intragranular cracking (Figure \ref{fig:CrackingExample-1}). The observed defect pattern supports the initial hypothesis that, even within macroscopically uniform areas of deposited alloys, thermal stress leads to microscopic cracking along the relatively brittle grain boundaries. Intergranular cracking is favored for a variety of reasons (e.g., lattice misfit, void/oxide formation) that are relatively composition-independent in these solid-solution alloys. Thus, intergranular cracking may vary because of differences in elasticity. In simple terms, stresses develop inside the microstructure due to anisotropy in the coefficient of thermal expansion and the lattice misorientation of grains. A lattice with low shear modulus can deform to accommodate this misfit stress, while a high-modulus lattice splits along boundaries with a high misorientation angle. We are interested in testing if the observed cracking is controlled by the grain boundary or by the lattice in this situation. If grain boundary character dominates this, cracking will be consistent across alloys; whereas if elasticity dominates, cracking will align with the Pugh ratio. In the next section, we resolve that hypothesis quantitatively. 

\subsection{Cracking - Pugh Ratio Correlation}

\noindent

\begin{figure}[h]
    \centering
    \includegraphics[width=1\textwidth]{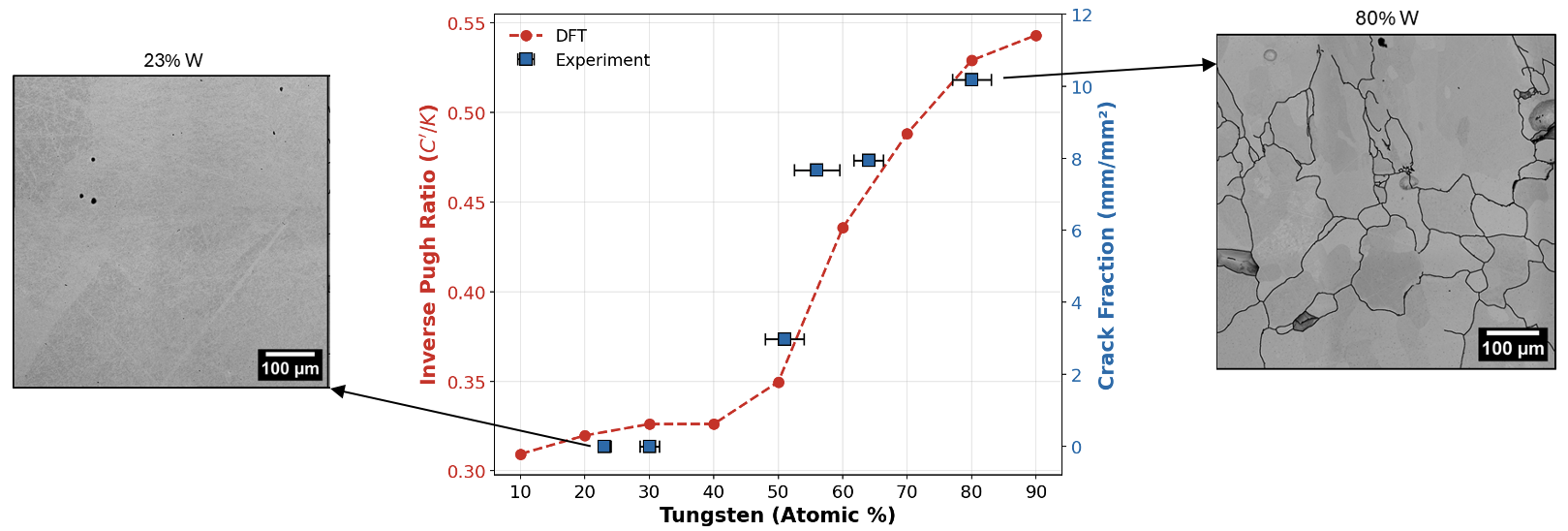}
    \caption{Correlation between Pugh ratio and crack fraction as a function of the tungsten fraction of the alloys. Examples of cracking in equally-sized micrographs from samples with minimum (23\%) and maximum (80\%) tungsten; cracks have been artificially traced after skeletonization for visual clarity.}
    \label{fig:pugh_crack_comparison_sqs}
\end{figure}

Figure \ref{fig:pugh_crack_comparison_sqs} presents the average crack fractions of printed W–Ta–Nb alloys measured within representative areas (as discussed in the supplemental materials). Error bars indicate the standard deviation of tungsten composition (as measured by EDS). Alongside the crack fraction is the \textit{inverse} of the Pugh ratio. The inverse Pugh ratio was used to more easily show the correlation with crack fraction. When the ductility is high, both the inverse Pugh ratio and the crack fractions are low. Consistent with the inverse Pugh ratio, alloys with higher tungsten content exhibit greater crack fractions, while those with lower tungsten content show significantly reduced cracking.  Overall, Figure \ref{fig:pugh_crack_comparison_sqs} demonstrates a reasonable correlation between the inverse Pugh ratio and the crack fraction, supporting the conclusion that cracking behavior in AM RMPEA's is directly related to elastic properties, with additional dependence on processing conditions and defects. This suggests that the inverse Pugh ratio can serve as a computational proxy for cracking behavior in additively manufactured tungsten-based refractory alloys. Notably, the \ce{W20Ta70Nb10} and \ce{W30Ta60Nb10} alloys exhibit no visible intergranular cracking, making them strong candidates for future studies to eliminate processing-related defects and examine mechanical properties.

\section{Conclusion}

Tungsten has exceptional thermal and chemical properties, which make it ideal for extreme environments applications. However, the low ductility of tungsten precludes the use of additive manufacturing, which is essential to the production of complex parts. Rather than screen testable alloys based on strength, this work advanced the ductility of high melting-point alloys. We established the Pugh ratio (K/$C^{\prime}$) as a criterion for intrinsic ductility by developing several novel conclusions:
\begin{itemize}
    \item The UMA machine learning interatomic potential was used to screen the W-Ta-Nb alloy compositions and identify the alloys with 10\% Nb to be optimal on the Pugh ratio and melting point.
    \item DFT was used to calculate the Pugh ratios of the 10\%Nb alloys more accurately and develop an electronic explanation of the trend in the Pugh ratio. 
    \item The tetragonal shear ($C^{\prime}$) was used instead of the average shear modulus ($G_{\text{VRH}}$) to calculate the Pugh ratio because it was shown to correlate better with the cracking mechanism of tungsten alloys.
    \item In printed W$_{\text{x}}$Ta$_{90-\text{x}}$Nb$_{10}$ alloys, macroscopic defects were predominantly caused by lack of fusion, porosity, and unmelted particles, while the primary microscopic defect was intergranular cracking.
    \item The ratio of intergranular crack length to image area shows a reasonable agreement with the inverse Pugh ratio in the examined alloys. Notably, the \ce{W20Ta70Nb10} and \ce{W30Ta60Nb10} alloys exhibit no visible intergranular cracking, making them strong candidates for future mechanical property studies.
    
\end{itemize}

These results provide a strong foundation for the further exploration of printable RMPEA compositions. Using Pugh ratio as a screening criterion, our alloy search can be more easily expanded to a broader range of elements and include quaternary and penternary compositions. Future work will seek to increase the strength of the \ce{W20Ta70Nb10} and \ce{W30Ta60Nb10} alloys by eliminating processing defects and incorporating an oxide dispersion.

\clearpage
\section{CRediT author statement}
\textbf{Kareem Abdelmaqsoud} Conceptualization, Methodology, Software, Formal Analysis, Writing - Original Draft, Writing - Review and Editing, Visualization
\textbf{Daniel Sinclair} Conceptualization, Methodology, Formal Analysis, Investigation, Writing - Original Draft, Writing - Review and Editing
\textbf{Amaranth Karra} Investigation, Writing - Review \& Editing
\textbf{S. Mohadeseh Taheri-Mousavi} Conceptualization, Project Administration, Funding Acquisition
\textbf{Michael Widom} Formal Analysis, Writing - Review \& Editing
\textbf{Bryan Webler} Conceptualization, Project Administration, Funding Acquisition, Writing - Review \& Editing
\textbf{John Kitchin} Conceptualization, Methodology, Project Administration, Funding Acquisition, Supervision

\section{Acknowledgement}
We acknowledge support from the Naval Nuclear Laboratory (NNL) under Award No. 1047622. Electronic structure and elasticity investigations by MW were supported by the Department of Energy Grant No. DE-SC0014506.
We acknowledge the contribution of the following groups and individuals to experimental processes:
Electron microscopy was performed in the Carnegie Mellon University Materials Characterization Facility.
Scott Kram: DED training, operation, and maintenance.
Ian McLachlan: EDM operation.

\clearpage
\bibliographystyle{elsarticle-num}
\bibliography{references}

\end{document}